\documentclass[cits]{PoS}
\usepackage[authoryear]{natbib}
\bibpunct{(}{)}{;}{a}{}{,}

\title{Stacking for Cosmic Magnetism with SKA Surveys}

\ShortTitle{Stacking for Cosmic Magnetism with SKA Surveys}

\author{\speaker{Jeroen M. Stil}\thanks{On behalf of the SKA Cosmic Magnetism Working Group}\\
        Department of Physics \& Astronomy, The University of Calgary\\
        E-mail: \email{jstil@ucalgary.ca}}

\author{Ben W. Keller\\
        Department of Physics \& Astronomy, MacMaster University\\
        E-mail: \email{kellerbw@mcmaster.ca}}

\abstract{Stacking polarized radio emission in SKA surveys provides statistical information on large samples that is not accessible otherwise due to limitations in sensitivity, source statistics in small fields, and averaging over frequency (including Faraday synthesis). Polarization is a special case because one obvious source of stacking targets is the Stokes I source catalog, possibly in combination with external catalogs, for example an SKA HI survey or a non-radio survey. We point out the significance of stacking subsamples selected by additional observable parameters to investigate relations that reveal more about the physics of the source. Applications of stacking polarization include, but are not limited to, obtaining in a statistical sense polarization information to the detection limit in total intensity, depolarization as a function of cosmic time at consistent source-frame wavelengths, magnetic field properties in objects with a low radio luminosity such as dwarf and low-surface-brightness galaxies, and investigating potential correlations of observable parameters with the average magnetic field direction in a sample. We also point out the potential use of stacking in validating the polarization calibration of a survey. While stacking is flexible in terms of survey definition, we discuss optimal survey parameters for the science experiments presented, as well as computing and archiving requirements. }

\FullConference{Advancing Astrophysics with the Square Kilometre Array,\\
		June 8-13, 2014\\
		Giardini Naxos, Sicily, Italy }

\newcommand{\skipthis}[1]{}

\newcommand\apj{ApJ}
\newcommand\aj{AJ}
\newcommand\mnras{MNRAS}

\begin{document}

\section{Introduction}

One of the areas of strength for the Square Kilometre Array (SKA) on the evolution of galaxies and their magnetic fields is in the analysis of very large samples over a wide range of redshift. Stacking is a statistical analysis of the emission of sources that are too faint to be detected individually in the survey, whose position is known from another survey. Stacking provides a flux density representative for a carefully selected sample of sources. The astrophysical interpretation depends on the selection criteria of the input catalog. A particularly useful application is to compare subsamples distinguished by an observable parameter, e.g. inclination, radio spectral index or flux at another wavelength, and morphological or spectroscopic classification. Interpretation of polarization stacking is more straightforward if one can further divide the sample into narrow bins of total flux density from the Stokes I catalog. 

Stacking radio polarization \citep{stil2014} offers three significant benefits. The first is that it allows investigation of the polarization of radio sources to the detection limit in total intensity, even though the polarized signal is usually only a few percent of the total flux density. The second is that stacking polarization as a function of flux density allows a uniform investigation of the polarization of radio sources without applying a detection threshold in polarized intensity. The third benefit of stacking polarization is that it provides an opportunity to study polarization of sources as a function of frequency with high sensitivity without the need to average over all observed frequencies. This is an important consideration for investigation of wavelength-dependent effects such as Faraday rotation and depolarization as a function of redshift.

\begin{figure}
\center
\resizebox{12cm}{!}{\includegraphics[angle=-90]{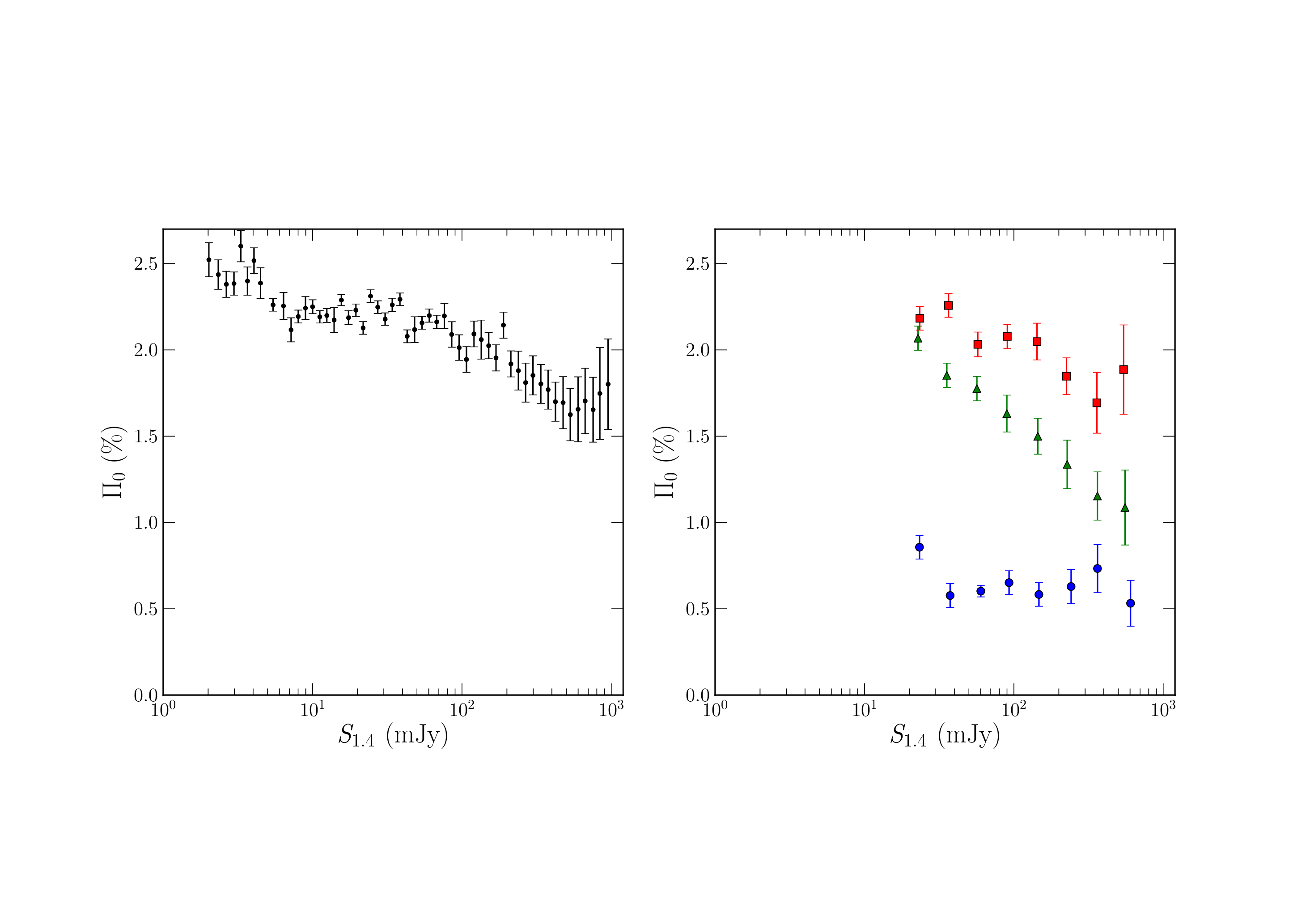}}
\caption{ Fractional polarization of radio sources at 1.4 GHz as a function of flux density from stacking NVSS \citep{condon1998} polarized intensity. Left: results from \cite{stil2014} using all NVSS sources. Right: preliminary results stacking sources as a function of spectral index in the area of overlap between the Westerbork Northern Sky Survey \citep[WENSS, 325 MHz][]{rengelink1997} and the NVSS. Blue circles: $\alpha > -0.3$, green triangles: $-0.75 < \alpha < -0.3$, red squares: $\alpha < -0.75$. The flux density range is limited on the high end because of the samples are smaller, and on the low end by the quality of the spectral indices of flat spectrum sources.
\label{PI_vs_I-fig}}
\end{figure}

Figure~\ref{PI_vs_I-fig} illustrates some of these advantages from stacking the NRAO VLA Sky Survey \citep[NVSS,][]{condon1998}. The left panel in Figure~\ref{PI_vs_I-fig} shows fractional polarization versus flux density for all NVSS sources. Each data point in this figure represents a stacking experiment and a sequence of Monte-Carlo realizations of the stack to correct for polarization bias and determine the error bars. Investigation of median fractional polarization based on direct detection of polarization of NVSS sources is limited to $S_{1.4} \gtrsim 80\ \rm mJy$ \citep{mesa2002,tucci2004}. While deep surveys can detect polarization in fainter sources, detecting the gradual change in fractional polarization in Figure~\ref{PI_vs_I-fig} requires sample sizes for which current deep fields are too small. The noise in the median image of the faintest bin is $1 \ \mu\rm Jy$, sufficient to detect polarization in the faintest sources in the NVSS catalog without concern for confusion. 

Splitting up the sample also by spectral index reveals how the dependence of polarization on flux density is related to sources with intermediate spectral index (Figure~\ref{PI_vs_I-fig}, right). A difference in median fractional polarization between bright flat and steep spectrum sources was reported by \cite{mesa2002} and \cite{tucci2004}, but the trends in Figure~\ref{PI_vs_I-fig} can only be studied through stacking at this time. Does the trend for for sources with intermediate spectral index continue at lower flux density? What model of radio source populations can reproduce these trends? This example illustrates how subdivision of the input catalog provides additional information by revealing a correlation with an observable parameter. The flux density limit in the right panel of Figure~\ref{PI_vs_I-fig} is set by uncertainty in the spectral index used to select the subsamples. We subdivided the sample at the modal spectral index $-0.75$ so that noise broadening of the spectral index distribution of faint sources does not affect the number of steep-spectrum sources in the intermediate spectral index range. 

This example illustrates how significant progress can be made using only information from the Stokes I source catalog of the same SKA survey. Such experiments are by definition free from confusion or source blending, and the astrophysical interpretation of the result in terms of fractional polarization is straightforward by subdividing the sample in bins of total flux density before stacking. If additional information is available, possibly only for some part of the survey area, it becomes possible to do additional stacking experiments that provide further insight in the physics of the polarized emission by revealing correlations with observable parameters, or even extend the investigation beyond the completeness limit of the Stokes I source catalog.

\section{Stacking Applications for Cosmic Magnetism}
\label{application-sec}

We do not aim to give an exhaustive description of all possible stacking experiments that can be done with an SKA polarization survey. The few applications mentioned here serve to illustrate how this analysis can advance science with the SKA, supplementing results obtained with other techniques.

\subsection{Fractional polarization as a function of redshift}

Faraday Rotation Measure (RM) Synthesis \citep{brentjens2005} averages polarization across the observed frequency range to optimize sensitivity and resolution in Faraday depth. Faraday rotation and depolarization in a source at redshift $z$ occur at the source frame wavelength $\lambda_{\rm src} = \lambda_{\rm obs}/(1+z)$. For observations related to the evolution of cosmic magnetism it is desirable to compare, for example, depolarization in similar sources at a consistent set of $\lambda_{\rm src}$, to have a model-independent reference for the no-evolution scenario. For Faraday RM Synthesis, the ability to track the same range in $\lambda_{\rm src}^2$ as a function of redshift requires a trade-off with sensitivity as a significant part of the observed frequency range is used to trace the same range in $\lambda_{\rm src}^2$ as a function of redshift. The added value of stacking in narrow frequency ranges is best illustrated by an example.

\begin{figure}
\center
\resizebox{9cm}{!}{\includegraphics[angle=0]{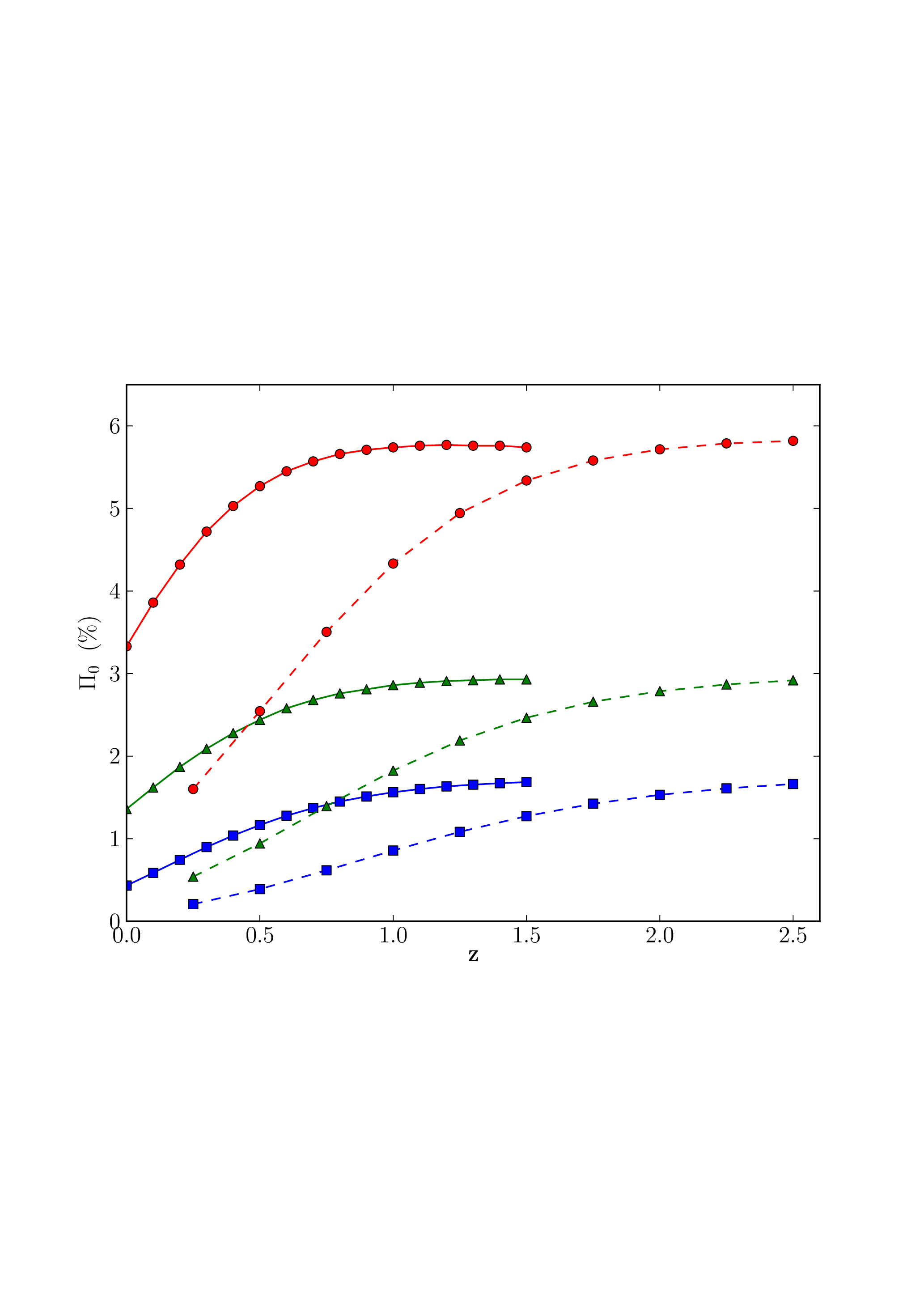}}
\caption{ Fractional polarization derived from the peak of the Faraday depth spectrum from Faraday synthesis of models of the integrated polarization of galaxies in SKA1-SUR band 3 (1.5 - 4.0 GHz; continuous curve) and SKA1-MID band 2 (0.95 - 1.76 GHz; dashed curve). Galaxy models were evaluated for inclination $60^\circ$, and ratio of random to regular magnetic field strength 2.0 (red circles), 3.0 (green triangles), and 4.0 (blue squares).  
\label{RM_syn_z-fig}}
\end{figure}

Figure~\ref{RM_syn_z-fig} shows fractional polarization from Faraday RM Synthesis of models of unresolved galaxies from \cite{stil2009} in two bands: SKA1-SUR band 3 and SKA1-MID band 2 \citep{dewdney2013}. Star forming galaxies make up a majority of the faint radio source population observable with the SKA, but the argument applies to any source type that is affected by depolarization. Figure~\ref{RM_syn_z-fig} shows a significanct increase of polarization with redshift for the same galaxy, as the observed frequency band shifts to shorter wavelengths in the source frame. The advantage of polarization stacking at consistent $\lambda_{\rm src}$ is its model-independent no-evolution case.

Stacking polarization averaged over narrow ranges of $\lambda_{\rm obs}$ (1 or more channels) for samples of galaxies as a function of redshift will allow us to derive fractional polarization as a function of $\lambda_{\rm src}$ to the extent allowed by the frequency range of the survey. SKA1-SUR band 3 for lower redshifts and SKA1-MID band 2 for higher redshifts (and higher sensitivity) are well positioned for this experiment. While Faraday synthesis of these galaxies provides a band-averaged fractional polarization, stacking will benefit from less depolarization at the short-wavelength end in the source frame. Any investigation as a function of redshift requires identification of optical counterparts with known redshifts, as well as rejection of galaxies with an AGN, for example through the infrared-radio correlation. Samples larger than $10^4$ per stack are required to derive the median polarization to the detection limit in total intensity. In practice, he analysis will be limited by the size and completeness of the optical target catalogue.

Stacking samples with known redshifts allows us to use the broad-band capability of the SKA to explore depolarization in a way that supplements information from Faraday RM Synthesis. This analysis requires cubes of Stokes $I$, $Q$, and $U$ to be archived. In Section~\ref{survey-sec} we address reduction of data volume by channel averaging without compromising the results.

\subsection{Probing deeper into the luminosity function}

There are various reasons to investigate the polarization properties of less luminous AGN in comparison with higher luminosity AGN. The FRI/FRII morphological distinction is related to luminosity \citep{fanaroff1974}. The different relative brightness of jets, lobes and hot spots suggests a physical background for the observed higher fractional polarization of faint AGNs.  Since part of the radio emission of these sources is beamed, fractional polarization may also be related to (isotropic) luminosity through the orientation of the source axis. Evolution and redshift-dependent depolarization may also play a role because the redshift range at any flux density is large due to the wide range in luminosity of AGN powered radio sources. 

Polarization of low-surface-brightness galaxies and dwarf galaxies, each with a lower radio to optical luminosity ratio than regular spiral galaxies, includes galaxies with less shear and a lower star formation rate per unit mass. These are important parameters in dynamo theory through the amount of shear and injection of turbulent energy through stellar feedback (see the chapter on nearby galaxies by R. Beck for a discussion). Using an HI selected sample that includes sub-selection on the ratio of HI mass to stellar mass, shape of the line profile (dynamics, shear) and inclination we can address the question whether magnetic fields in quiescent galaxies with a low surface brightness are different from those in high-surface brightness galaxies with strong continuum emission. An HI selected sample has complete redshift information by definition, although this work is limited in redshift by the depth of the HI survey. In this case, stacking polarization extends magnetic field investigations to galaxies that occupy a part of parameter space that is not accessible by other means.

\subsection{Stacking Stokes Q and U}

At first sight it may not appear useful to consider Stokes $Q$ and $U$ individually for stacking. However, if a potential predictor for polarization angle can be identified, its correlation with polarization angle can be tested by stacking samples based on the value of the predictor, or the sample can be rotated by a parallactic angle based on the predictor before stacking. A simple but effective significance test is randomizing the orientation of the sample. If the sample is also stacked in polarized intensity, then the median $Q$ and $U$ may be compared with the median polarization of the sample in order to model the degree of correlation with polarization angle. Possible predictors include optical polarization angle, optical minor axis for disk galaxies, or orientation of the jet. 

If a predictor for polarization angle is identified, one can stack the sample as a function of frequency, to investigate if the predictor for polarization angle is equally valid at longer wavelengths where Faraday rotation is stronger. If the sample is mostly Faraday thin at the highest frequencies, but experiences significant and diverse Faraday rotation at lower frequencies, this should be observable as the median stacked $Q$ and $U$ converge to zero as frequency decreases. Combining this with the behavior of the median fractional polarization in the same frequency range allows one to distinguish between Faraday rotation and depolarization. As for polarized intensity, extending this analysis to the faintest sources detectable in total intensity should be straightforward, depending on the availability of a predictor for polarization angle. Variation of resolution across the band should be considered, and can be dealt with by integrating $Q$ and $U$ over an aperture before stacking.

\subsection{Stacking in relation to commissioning and data verification}

Stacking requires a sound understanding of the data, including the statistics of the noise and systematics. A statistical analysis of many sources may reveal subtle features of the data that are difficult to identify otherwise. Examples are preferred polarization angles in the NVSS survey \citep{battye2008}, the effect of clean bias in stacking quasars in the FIRST survey \citep{white2007}, and bandwidth depolarization in the NVSS \citep{stil2007}. 

Direction-dependent quality control for polarization calibration is challenging because of the limited number of bright polarized sources. A stacking analysis of samples selected by their position in the field (location with respect to the beam center or location on a phase array feed) can be used as part of the verification of polarization purity of a commissioning survey. This kind of analysis requires the necessary meta data about field centers to be available at least for expert users.

\section{Optimal surveys for stacking experiments}

\label{survey-sec}

Most surveys can be considered for stacking experiments. The sensitivity of the stacking experiment shown in Figure~\ref{PI_vs_I-fig} is limited by the depth of the Stokes I source catalog, so a reduction in sensitivity will reduce the depth of a stacking experiment accordingly. There are good reasons to stack both wide and deep surveys. Wide surveys tend to be more effective because of the much larger number of target sources, while the actual sensitivity for deep surveys may be limited by samples size rather than noise \citep[see ][for a discussion]{stil2014}. However, deep surveys target areas of the sky with more ancillary data and usually have higher angular resolution. These allow more detailed sample selection and differentiation. A deep survey with relatively constant noise except for the edges of the survey area is more suitable for stacking polarization. Sample size and variation of the noise are included in Monte-Carlo simulations that correct for polarization bias and estimate error bars.  

Confusion is the most significant factor in deciding the suitability of a survey for a particular stacking experiment. For example, it is not possible to investigate the polarization of SDSS quasars by stacking the NVSS survey, because the density of the targets is so high that one is forced to stack well below the confusion limit of the NVSS. While confusion is a problem stacking total intensity, it is prohibitive for stacking polarization because of blending of sources with different polarization angle. This is only an issue for samples selected at other wavelengths, because stacking based on the Stokes I source catalog is by definition free from confusion. The monochromatic imaging performance of SKA1-SUR and SKA1-MID allow imaging at resolution between $1''$ and $5''$ that will allow us to avoid confusion in stacking optically selected samples \citep{braun2013}.  

Higher angular resolution results in a lower tolerance for position errors in the input catalog, and some targets may be resolved. Targets selected from an X-ray or ultraviolet source catalog can have substantial position errors due to limited photon statistics, and the X-ray emission may be offset from the radio position. All of these complicate the interpretation of the stacking result as a representative flux density for the sample. As mentioned before, these issues can be mitigated by integrating Stokes $I$, $Q$, and $U$ over an aperture centered on the target position by the stacking software. This feature requires more flexibility of the stacking code and some more computing resources. 

\begin{figure}
\center
\resizebox{7cm}{!}{\includegraphics[angle=0]{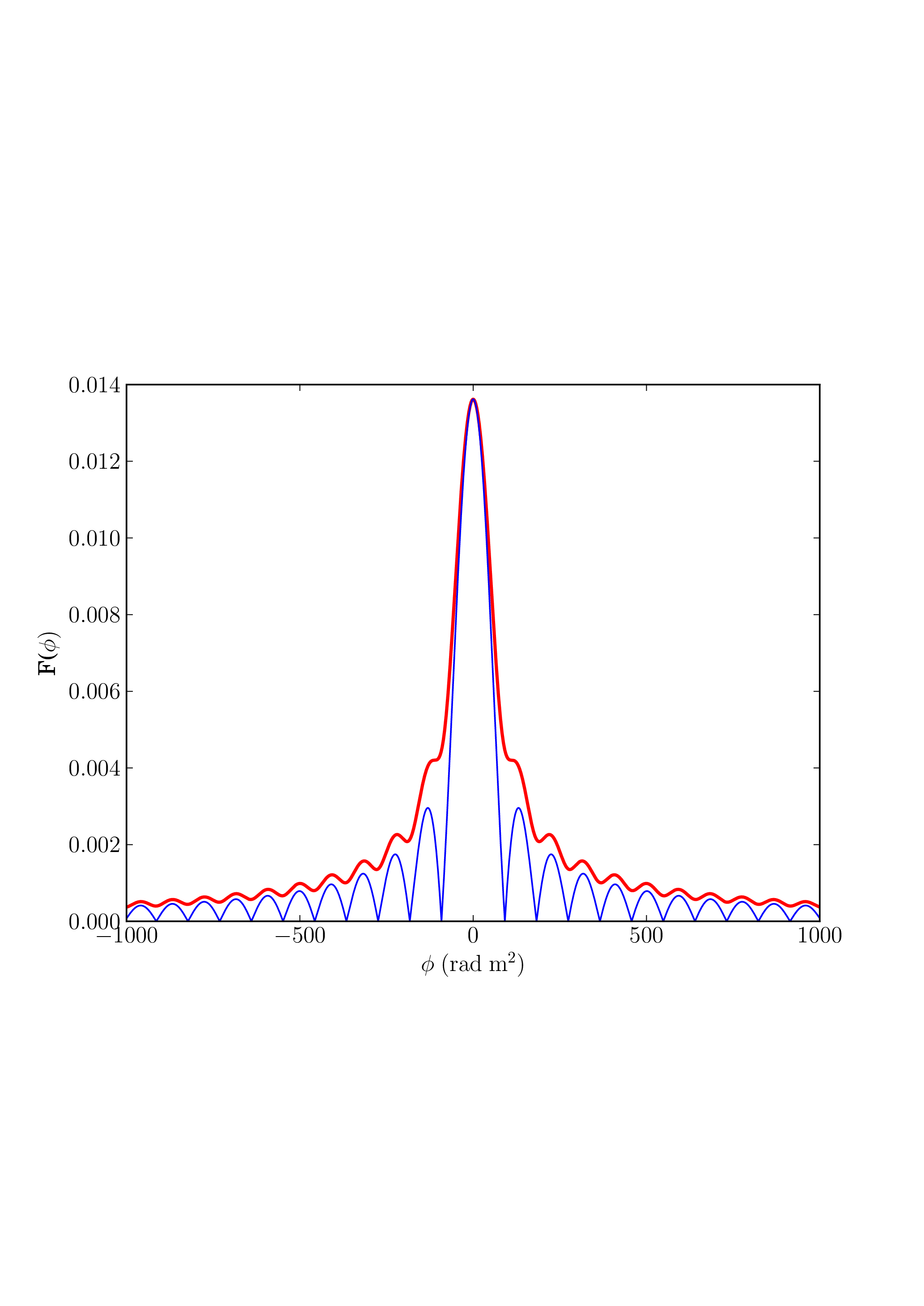}}
\resizebox{7cm}{!}{\includegraphics[angle=0]{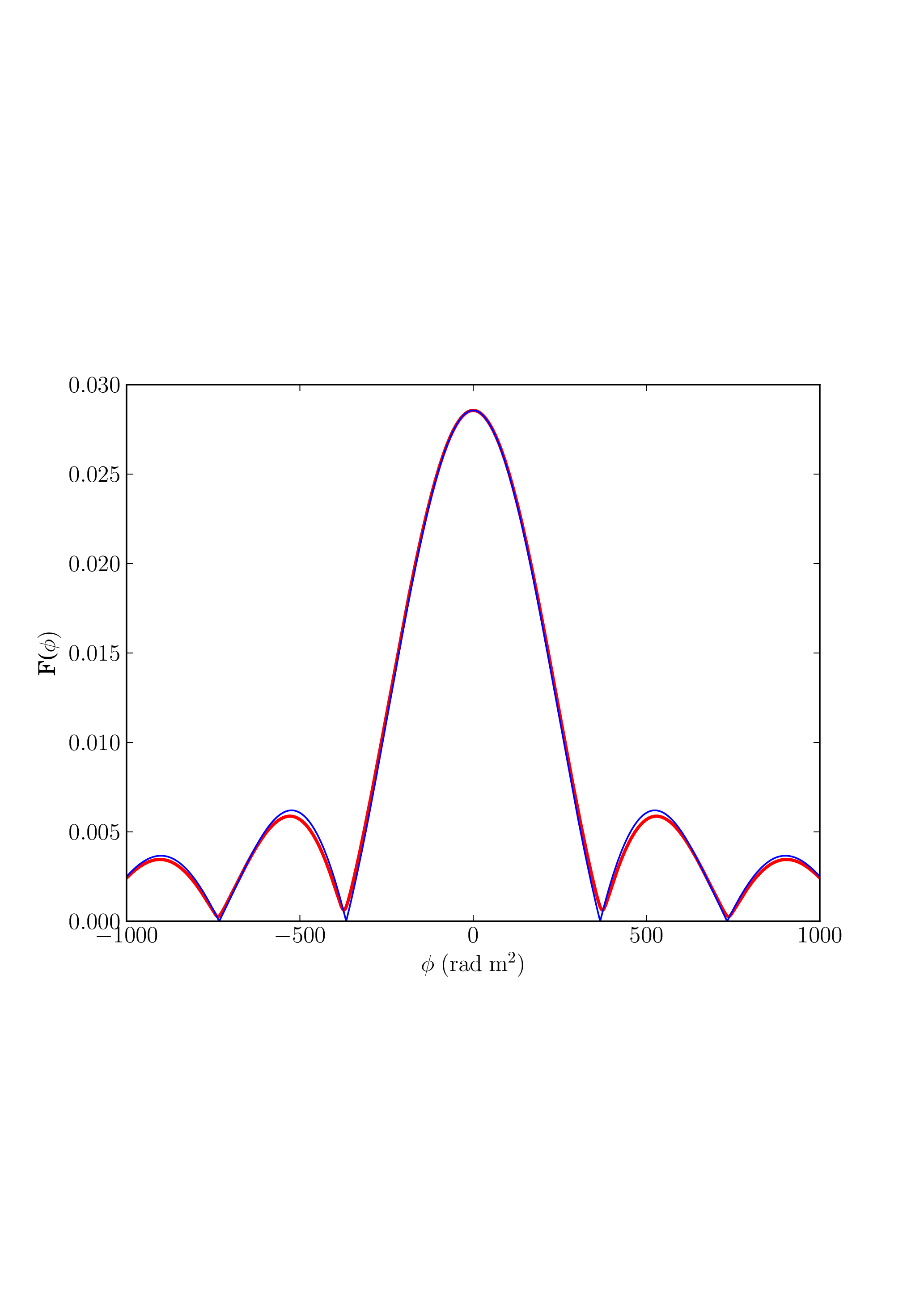}}
\vskip 1.8cm
\caption{ Model Faraday depth spectra (red) and associated RMTF (blue) of the same galaxy at two different redshifts, observed in SKA1-SUR band 3. Left: $z = 0$, right: $z = 1$. At low redshift, the Faraday depth spectrum is clearly resolved, revealing Faraday rotation of diffuse emission in the disk. If the same galaxy at $z = 1$ is observed in the same frequency band, the resolution in Faraday depth is much reduced, allowing, in principle, stacking of Faraday depth spectra. Averaging channels before stacking is equivalent but computationally more efficient. \label{RM_syn_spec-fig}
}
\end{figure}

Channel averaging before stacking can increase the sensitivity before stacking and reduce the computational footprint and data volume of the stacking analysis. It is equivalent to Faraday synthesis under the assumption that Faraday rotation is negligible over the frequency range that is averaged. The question how much channel averaging can be done before stacking is therefore analogous to that of alignment errors in stacking: as long as the errors remain significantly smaller than the resolution of the survey, their effect is small. Potential issues with Faraday thick sources are no different from issues encountered when stacking resolved HI line profiles of galaxies. If some of the targets experience significant Faraday rotation over the frequency range that is averaged, then these sources will appear depolarized after channel averaging $Q$ and $U$. Any difference between stacking channel averaged polarized intensity and polarized intensity calculated from channel averaged $Q$ and $U$ can be used as a test to decide if this is significant.

The resolution in Faraday depth in an object at redshift $z$ using only data from wavelength $\lambda_{\rm obs,1}$ to $\lambda_{\rm obs,2}$ is \citep{brentjens2005}
\begin{equation}
\Delta \phi_z = {2 \sqrt{3} \over \lambda_{\rm obs,2}^2 - \lambda_{\rm obs,1}^2}(1+z)^2,
\label{FD_res-eq}
\end{equation}
Figure~\ref{RM_syn_spec-fig} illustrates the effect of redshift on the Faraday depth spectrum of two identical objects observed in the same frequency range. For averaging channels one should always consider at least the Galactic foreground at $z=0$ and the range of Faraday depths in the sample. At high Galactic latitude the foreground rotation may be estimated and removed before averaging with accuracy of a few rad m$^{-2}$, with a computational overhead that is still small compared to Faraday synthesis. Compression of the $\lambda^2$ range and shift to shorter wavelength allows for a greater tolerance for the unknown Faraday depth range of the source if a at least a lower limit for the redshift of the sample is available.Equation~\ref{FD_res-eq} can thus be inverted to derive the wavelength range that can be averaged before stacking polarization.

\section{Archive and Computing Resources}

\subsection{Stacking Computational Resources}
Generally, polarization stacking is computationally efficient, and amenable to
parallelized implementation.  The primary direct cost for median stacking is the
selection of median values for each pixel.  For stacked images $m$ pixels
across, this naturally means $m^2$ selections must be performed. Modern
selection algorithms, such as \textit{quickselect}, as used by \cite{stil2014}, have 
linear scaling to the number of sources in the average
case.  Thus, for a stack of $N$ sources with $m\times m$ images, the
computational complexity is $\mathcal{O}$ $(m^2 N)$, with the same spatial
complexity (quickselect can be sorted in-place).  However, the simple task of
generating median values from a generated stack is not the only cost associated
with stacking analysis.  For example, if one wishes to use oversampled
source alignment prior to generating a stack, each of $N$ images must be
interpolated to a higher resolution.  This task can be accomplished in parallel.

The largest practical value of N depends on polarization purity of the survey, as well as the range at which $\sqrt{N}$ improvement of the noise can be achieved. Assuming 0.1\% polarization purity, a practical limit at the detection threshold in Stokes I is $N \sim 10^6$ per stack of a single $\sim$0.1 dex flux density bin. Provided that $\sqrt{N}$ improvement up to $N = 10^6$ can be achieved, Stacking an all sky-survey will be limited by polarization purity, whereas a deep field will be limited by sample size. The scientific experiment will involve orders of magnitude more sources as flux density and parameters such as spectral index are included. Stack rare objects will require a complete sky survey to collect sufficiently large samples. 

\subsection{Bias Correction Computational Costs}
As \cite{stil2014} showed, bias correction for a polarized intensity stack is non-trivial.  With Monte-Carlo realizations, a significant amount of CPU time is required to generate an estimate for the true stacked polarization.  For a typical NVSS stack, roughly 5 CPU hours (on an AMD Opteron 2218 CPU) was required.  However, little time has been spent optimizing our Monte-Carlo method, and this number can likely be reduced by an order of magnitude or more.  Beyond this, the reduction for a large number of stacks containing many sources is extremely amenable to parallelization.  A parallel implementation should be capable of scaling to many hundreds to thousands of CPUs if necessary.

\subsection{Storage and Archive Requirements}

While stacking can be used to push well beyond the sensitivity of a survey, a significant new application of this technique to wide-band radio data is to the statistics of samples as a function of frequency. These are not limited to stacking polarization \citep[e.g.][]{stil2014}. These applications depend on archiving image cubes, as the band-averaged images do not contain the necessary information, although reduction of the archived data volume by an order of magnitude or more can be achieved by averaging piecewise over channels. Stacking polarization invariably requires stacking total intensity in the same frequency range. We expect cubes of total intensity to be made for the purpose of other magnetism science cases.

Working from existing survey images, stacking analysis requires negligible additional storage resources. With double-precision pixel values, each $m\times m$ stacked image requires a meager $16 m^2$ bytes per stacked sample. Stacking analysis on-the-fly (a stacking survey) offers the opportunity to greatly reduce the storage costs associated with storing massive amounts of survey data.  By discarding data once a stack has been generated, the storage needed for generating a stack of $N$ sources can be reduced by a factor of $N$.  For large stacks, this can mean a reduction in storage requirements by many orders of magnitude. We stress that scaling this analysis to an SKA survey does not require that $N$ scales proportional to the size of the source catalog because of subselection by flux density and spectral index. Memory requirements per CPU are therefore not expected to be a limiting factor.

\bibliographystyle{apj}

\end{document}